\documentclass[12pt]{iopart}
\usepackage{graphicx}
\usepackage{cite}
\eqnobysec

\begin{document}

\title[Two interacting polymer chains on fractals]
{Critical behavior of  interacting two-polymer system in a fractal
solvent: an exact renormalization group approach}

\author{I \v Zivi\'c\dag, S Elezovi\'c-Had\v zi\'c\ddag~and S Milo\v sevi\'c\ddag}

\address{ \dag Faculty of Natural Sciences and Mathematics,
University of Kragujevac, 34000 Kragujevac, Serbia}
\address{\ddag Faculty of Physics,
University of Belgrade, P.O.Box 368, 11001 Belgrade, Serbia}

\eads{\mailto{ivanz@kg.ac.yu}, \mailto{suki@ff.bg.ac.yu},
\mailto{emilosev@etf.bg.ac.yu}}

\begin{abstract}
We study the polymer system consisting of  two  polymer chains
situated in a fractal container that belongs to the
three--dimensional Sierpinski Gasket (3D SG) family of fractals.
Each 3D SG fractal has four fractal impenetrable 2D surfaces,
which are, in fact, 2D SG fractals. The two-polymer system is
modelled by two interacting self-avoiding walks (SAWs), one of
them representing a 3D floating polymer, while the other
corresponds to a chain adhered to one of the four 2D SG
boundaries. We assume that the studied system is immersed in a
poor solvent inducing the intra-chain interactions. For the
inter-chain interactions we propose two models: in the first model
(ASAWs) the SAW chains are mutually avoiding, whereas in the
second model (CSAWs) chains can cross each other. By applying an
exact Renormalization Group (RG) method, we establish the relevant
phase diagrams for $b=2,3$ and $b=4$ members of the 3D SG fractal
family for the model with avoiding SAWs, and for $b=2$ and $b=3$
fractals for the model with crossing SAWs. Also, at the
appropriate transition fixed points we calculate the contact
critical exponents, associated with the number of contacts between
monomers of different chains. Throughout the paper we compare
results obtained for the two models and discuss the impact of the
topology of the underlying lattices on emerging phase diagrams.
\end{abstract}

\maketitle

\section{Introduction}
\label{uvod}

The self--avoiding walk (SAW) is   well-disposed as a standard
lattice model for a  flexible linear polymer chain in various
types of solvents \cite{vc}. In this model, the monomers that
comprise a polymer chain are related to the steps of a random walk
that must not contain self-intersections, while the surrounding
solvent is represented by an underlying lattice. In a good
solvent, with each step of the SAW we associate the same weight
factor $x$, while in a poor solvent, when two non-consecutive
monomers of a polymer chain become nearest neighbors, we introduce
the additional statistical factor $u$, which corresponds to the
intra-chain energy $\epsilon_u<0$. Even though an isolated polymer
chain is difficult to observe experimentally, numerous studies of
the single-chain statistics have been upheld as a requisite step
towards understanding the statistics of many-chain systems. A
natural extension of a single polymer concept is a model of two
interacting linear polymers, which may be relevant to perceive
behavior of multicomponent polymer solutions \cite{palissetto}. To
study the critical properties of the two-polymer system we shall
apply the following two models: The first is the model of two
mutually avoiding self-avoiding walks (ASAWs), whose paths on a
lattice cannot cross each other, and the second is the model of
two mutually crossing self-avoiding walks (CSAWs), that is, the
case of the two SAWs whose paths can intersect each other. Various
types of models with two avoiding SAWs  have been successfully
applied in the studies of phase transition of diblock copolymers
\cite{stella1,stella2}, as well as in the studies of unzipping
double-stranded  DNA molecules \cite{dna1,dna2,dna3,dna4}. On the
other hand, the model with two crossing SAWs was applied  for
studying the collapse transition of two-chain interacting system
on three- and four-simplex  lattice \cite{ks93,kspa1}, and
Euclidean lattices \cite{leoni}, as well as to study two randomly
interacting directed polymers on diamond hierarchical lattice
\cite{bata,haddad2}.

In this paper we apply both  ASAWs and CSAWs model to study the
two-polymer system that displays both intra- and inter-chain
interactions, on the three-dimensional (3D) fractal lattices,
which belong to the Sierpinski gasket (SG) family of fractals. We
assume that one of the two polymers is a floating chain in the
bulk of a 3D SG fractal, while the other is a polymer chain that
stays affixed to one of the four boundary  surfaces (being
actually 2D SG fractals)
 \cite{ZivicJSTAT}.  In the ASAWs model we assume that two
SAWs are in contact when they approach each other at the distance
which is equal to a lattice constant, and in this situation we
ascribe the contributing contact energy $\epsilon_v$ to the total
model energy. Similarly, in the CSAWs model we assume  that each
crossing between two SAW paths corresponds to a contact of two
monomers that belong to different polymer chains, and therefore we
associate the contact energy $\epsilon_c$ with such a crossing.
Since in both models, one  of two polymers is adhered to one of the fractal boundary surfaces, and because its monomers  take effect of surface interacting points for the bulk floating polymer chain, the proposed  models may also
be of interest for the problem of surface-interacting polymer chain in  homogeneous  \cite{r1,r2,r3} and disordered media \cite{ustenko}.

The main goal of this study is to establish phase diagrams in the space of interaction parameters (which consists of the intra- and inter-chain
interaction energy parameters), for both models, as well as to calculate the
contact critical exponents that describe behavior of  numbers of
monomer-monomer contacts between two polymer chains.

This paper is organized as follows. In section~\ref{ASAWs} of the
paper, we first describe the 3D SG fractals for general scaling
parameter $b$, as well as the ASAWs model. Then, we present the
general framework of an exact renormalization group   method,
within the model, and elaborate on the phase diagrams,  obtained
for the fractals designated by $b=2,3$ and $b=4$. We also display
our findings for the contact exponents (associated with the number
of contacts between the two SAWs). The CSAWs model is described in
section~\ref{CSAWs}. Again, by applying an exact RG method, which
is in the latter case technically more complicated, phase diagrams
and contact exponents for $b=2$ and 3 SG fractals are obtained,
and discussed. Brief summary and the concomitant conclusion are
presented in section~\ref{sumiranje}. Explicit form of the RG
equations for particular fractals are given in appendices.

\section{The model of two evading self-avoiding walks}
\label{ASAWs}

In this section we are going to apply the renormalization group
(RG) method to the model of two mutually avoiding self-avoiding
walks  on the 3D SG family of fractals. First, we  give a summary
of the basic properties of these fractals. We start with recalling
the fact that each member of the 3D SG fractal family is labeled
by an integer $b\ge 2$ and can be constructed in stages. At the
first stage ($r=1$) of the construction there is a tetrahedron of
base $b$ that contains $b(b+1)(b+2)/6$ upward oriented unit
tetrahedrons. The subsequent fractal stages are constructed
recursively, so that the complete self-similar fractal lattice can
be obtained as the result of an infinite iterative process of
successive $(r\to r+1)$ enlarging the fractal structure $b$ times,
and replacing the smallest parts of enlarged structure with the
initial ($r=1$) structure. Fractal dimension $d_f$ of the 3D SG
fractal is equal to $d_f^{3D}={{\ln [{{b(b+1)(b+2)}/6}}]/{\ln
b}}$. Each of the four boundaries of the 3D SG fractal is itself a
2D SG fractal, with the fractal dimension
$d_f^{2D}=\ln[b(b+1)/2]/\ln b\>$.

In the terminology that applies to the SAW, we assign the weight
$x_3$ to a step of the SAW in the bulk (3D SG fractal), which
represents a floating polymer (we mark it by $P_3$), and the
weight $x_2$ to a step of the SAW performed on one of the fractal boundaries (2D SG
fractal), which represents a 2D surface-adhered  polymer (marked
by $P_2$), whose  monomers act as interacting counterparts for monomers of the 3D
polymer chain. To describe the intra-chain interaction of $P_3$
chain, we introduce the Boltzmann factor $u=e^{-\epsilon_u/k_BT}$,
where $\epsilon_u<0$ is the interaction  energy of  two
non-consecutive neighboring monomers of $P_3$.

\begin{figure}
\hskip4cm
\includegraphics[scale=0.4]{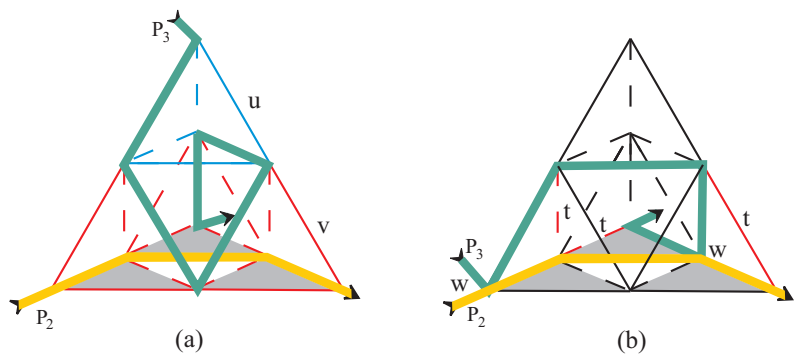}
\caption{The structure of the three-dimensional SG fractal, for
$b=2$, at the first stage of construction, with an example of the
bulk polymer chain ($P_3$) depicted by green  line and the
surface-adhered polymer chain ($P_2$) depicted by yellow line. The
shaded area represents the adhering surface (the two-dimensional
SG fractal). The intra-chain interactions $u$, for the $P_3$
polymer, are depicted by blue bonds.  In the ASAWs model (a) the
SAW paths, representing $P_3$ and $P_2$ polymers, cannot intersect
each other, and two SAWs interact when approach each other at a
distance which is equal to a lattice constant (red bonds, weighted
with $v$). On the other hand, in the case of CSAWs model (b), the
polymers $P_3$ and $P_2$ are cross-linked at the two sites, so
that each contact contributes the weight factor $w$, while the red
bonds (marked by $t$) correspond to the interactions between those
monomers which are nearest neighbors to the cross-linked points.
The two depicted examples  for  ASAWs (a) and CSAWs models (b),
contribute the weights
 $x_3^{5}x_2^{3}u^4v^{12}$ and $x_3^{4}x_2^{3}w^2t^3$, respectively.}
 \label{fig:interakcije}
\end{figure}

\begin{figure}
\hskip4cm
\includegraphics[scale=0.4]{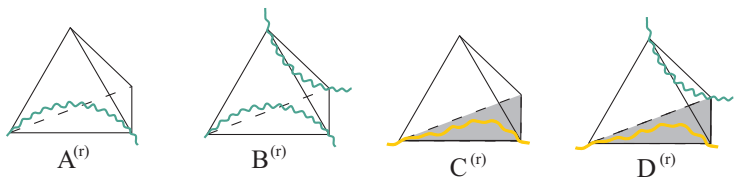}
\caption{Schematic depiction  of restricted generating functions
used in the description of all possible two-SAW configurations,
within the $r$-th stage of the 3D SG fractal structure, for ASAWs
model. The 3D floating chain is depicted by green line, while the
2D surface-adhered chain  is depicted by the yellow  one. The
interior details of the $r$-th stage fractal structure, as well as
details of the chains, are not shown (for the chains, they are
manifested by the wiggles of the SAW paths). The functions
$A^{(r)}$, $B^{(r)}$, and $C^{(r)}$, describe one-polymer
configurations (they are the same for both ASAWs and CSAWs
models), while the  function $D^{(r)}$  depicts the inter-chain
configurations of ASAWs model.} \label{fig:RGparametri}
\end{figure}

In ASAWs model the two SAWs, that represent polymer chains, must
not intersect each other. We assume that monomers,  belonging to
different chains, interact when they reach a distance which is
equal to a fractal lattice constant, and to a such mutual position
of $P_3$ and $P_2$ monomers we associate the weight factor
$v=e^{-\epsilon_v/k_BT}$ (see figure \ref{fig:interakcije}(a)),
where $\epsilon_v\leq0$ is the appropriate inter-chain interaction
energy.
To describe exactly all possible configurations of the two-chain
polymer system within the adopted  model, we need four restricted
partition functions $A^{(r)}$, $B^{(r)}$, $C^{(r)}$ and $D^{(r)}$,
which are defined as
\begin{eqnarray}
\fl A^{(r)}=\sum_{N_3,L} {\mathcal A}^{(r)}(N_3,L)  x_3^{N_3} u^L,\quad &&  B^{(r)}=\sum_{N_3,L} {\mathcal B}^{(r)}(N_3,L)  x_3^{N_3} u^L,\quad \nonumber\\
\fl C^{(r)}=\sum_{N_2} {\mathcal C}^{(r)}(N_2)  x_2^{N_2},\quad && D^{(r)}= \sum_{N_2,N_3,L,M}{\mathcal D}^{(r)}(N_2,N_3,L,M) x_2^{N_2}x_3^{N_3}u^Lv^M, 
\end{eqnarray}
where ${\mathcal A}^{(r)}$, ${\mathcal B}^{(r)}$, ${\mathcal C}^{(r)}$, and ${\mathcal D}^{(r)}$ represent the numbers of particular configurations, consisting of one or two SAW strands on the $r$-th fractal structure (see figure \ref{fig:RGparametri}). For instance, ${\mathcal D}^{(r)}(N_2,N_3,L,M)$ is the number of configurations consisting of $N_3$-step $P_3$ chain with $L$ pairs of non-consecutive nearest-neighbor monomers, and $N_2$-step $P_2$ chain, such that there are $M$ contacts between these two chains.
The recursive nature of the fractal construction implies the following recursion
relations for restricted partition functions
\begin{eqnarray}
 A'&=&\sum_{N_{A},N_{B}} a(N_{A},N_{B})\, A^{N_{A}}
B^{N_{B}}\,,
\label{eq:RGA}\\
 B'&=&\sum_{N_{A},N_{B}} b(N_{A},N_{B})\, A^{N_{A}}
B^{N_{B}}\,,
\label{eq:RGB}\\
 C'&=&\sum_{N_C} c(N_C)\, C^{N_C}\,,
\label{eq:RGC}\\
  D'&=&  \sum_{N_{A},N_{B},N_{C},N_{D}}
     d(N_{A},N_{B},N_{C},N_{D})\,
 A^{N_A}B^{N_B}C^{N_C}
 D^{N_{D}}
\>,
\label{eq:RGA4}
\end{eqnarray}
where we have used the prime symbol as a
superscripts for $(r+1)$-th restricted partition functions and no
indices for the $r$-th order partition functions.
These relations can be
considered as the RG equations for the problem under study, with
the  initial conditions
\begin{equation}\label{initalmodel2}
A^{(0)}=x_3\,,\quad B^{(0)}=x_3^2u^4\,,\quad C^{(0)}=x_2\,,\quad
\,D^{(0)}=x_3x_2v^4\,,
\end{equation}
which correspond to the unit tetrahedron\footnote{Such initial
conditions imply that a SAW can traverse unit tetrahedrons along
only one, or two nonconsecutive edges. This restriction of the
standard SAW model does not alter the critical behavior of the
system.}.

Equation (\ref{eq:RGC}), alone, describes a single SAW on 2D SG
 fractal, whereas (\ref{eq:RGA}) and (\ref{eq:RGB}) are RG equations
 for a single  SAW on 3D SG fractal. Critical properties of the SAW,
 based on the analysis of these equations, have been well established
 previously, and here we recall their basic properties relevant for
 the present work.

First, we describe the  behavior of a  single 2D SG chain. The RG
equation (\ref{eq:RGC}), for any $b$, has only one non-trivial
fixed point $C^*$, corresponding to the extended polymer phase
\cite{dhar78,EKM}, that is,  the 2D SG chain is always swollen,
and it cannot be in the compact phase. The corresponding
eigenvalue $\lambda_{\nu_2}$ of (\ref{eq:RGC}) is larger than 1,
and determines the value of the critical exponent $\nu_2=\ln b/\ln
\lambda_{\nu_2}$, that governs   the behavior of the mean
end-to-end distance of 2D SG chain $\langle R\rangle\sim {\langle
N_2\rangle}^ {\nu_2}$, where $\langle N_2\rangle$ is the average
number of 2D SG SAW steps.

In what follows we provide short summary of the results concerning
the critical behavior of a solitary 3D SG chain. Depending on the
value of the intra-chain interaction parameter $u$, a single 3D SG
chain can be found in three phases: extended chain (for
$u<u_\theta$), $\theta$-chain (when $u=u_\theta$) and globule
($u>u_\theta$). These phases (for arbitrary  $b$) are described by
the fixed points $(A_E,B_E)$, $(A_\theta,B_\theta)$ and
$(A_G,B_G)$, respectively \cite{DharVannimenus,Knezevic,EZM}. The
mean end-to-end distance $\langle R\rangle\sim {\langle
N_3\rangle}^ {\nu_3}$ of SAW on 3D SG fractal, is equal to
$\nu_3=\ln b/\ln \lambda_{\nu_3}$, where $\lambda_{\nu_3}$ is the
largest eigenvalue of the  linearized RG equations (\ref{eq:RGA})
and (\ref{eq:RGB}), at the corresponding fixed point. For each 3D
SG fractal, the following relationship
$\nu_3^E>\nu_3^\theta>\nu_3^G$, is valid, where $\nu_3^E$,
$\nu_3^\theta$ and $\nu_3^G$ are the end-to-end distance critical
exponents in extended, $\theta$ and globule phase, respectively.

The interacting configurations of $P_2$ and $P_3$ chains are described with the
 restricted partition function $D^{(r)}$. The mean number of contacts between
 $P_2$ and $P_3$, on the $r$-th stage of fractal construction, is equal to
 \begin{equation}\label{eq:srednjeMASAWs}
\langle M^{(r)}\rangle= {1\over D^{(r)}}\sum_{N_2,N_3,L,M}M{\mathcal D}^{(r)} x_2^{N_2}x_3^{N_3}u^L v^M =
 {v\over  D^{(r)}}
 \frac{\partial D^{(r)}}{\partial v}
 \>.
\end{equation}
On the other hand, taking into account the function dependance  $D^{(r+1)}=D^{(r+1)}(A^{(r)},B^{(r)},C^{(r)},D^{(r)})$,
and the fact that   $A^{(r)}$, $B^{(r)}$, and $C^{(r)}$ do not depend on the interaction parameter $v$, we have
\begin{equation}\label{dda}
\frac{\partial D^{(r+1)}}{\partial v}=\frac{\partial D^{(r+1)}}{\partial D^{(r)}}\frac{\partial D^{(r)}}{\partial v}\, ,
\end{equation}
from which follows that, in the vicinity of the transition fixed
point $(A^*, B^*, C^*, D^*)$ of the two-polymer system, the mean
number of contacts $\langle M^{(r)}\rangle$, for large $r$,
behaves as  $\langle M^{(r)}\rangle\sim \lambda_{D}^r
\label{eq:lambdaD}$, where
\begin{equation}\label{svvrednost2}
\lambda_{D}={\left(\frac{\partial D^{(r+1)}}{\partial D^{(r)}}\right)}^*\>,
\end{equation}
is relevant eigenvalue of RG equation (\ref{eq:RGA4}), calculated at the transition fixed point.
Knowing that $\langle {N_3^{(r)}}\rangle\sim \lambda_{\nu_3}^r$,
one obtains $\ln \langle M^{(r)}\rangle/{\ln \langle N_3^{(r)}\rangle}\sim {\ln\lambda_{D}}/{\ln\lambda_{\nu_3}}$, {\em i.e.} the following scaling relation is satisfied
\begin{equation}\label{asawfi}
 \langle M^{(r)}\rangle\sim \langle N_3^{(r)}\rangle^{\phi}\>,
\end{equation}
 where
\begin{equation}
\phi=\frac{{\ln\lambda_{D}}}{\ln\lambda_{\nu_3}}\, ,  \label{eq:skaliranje}
\end{equation}
is so-called contact critical exponent.

\begin{figure}
\begin{center}\includegraphics[scale=0.5]{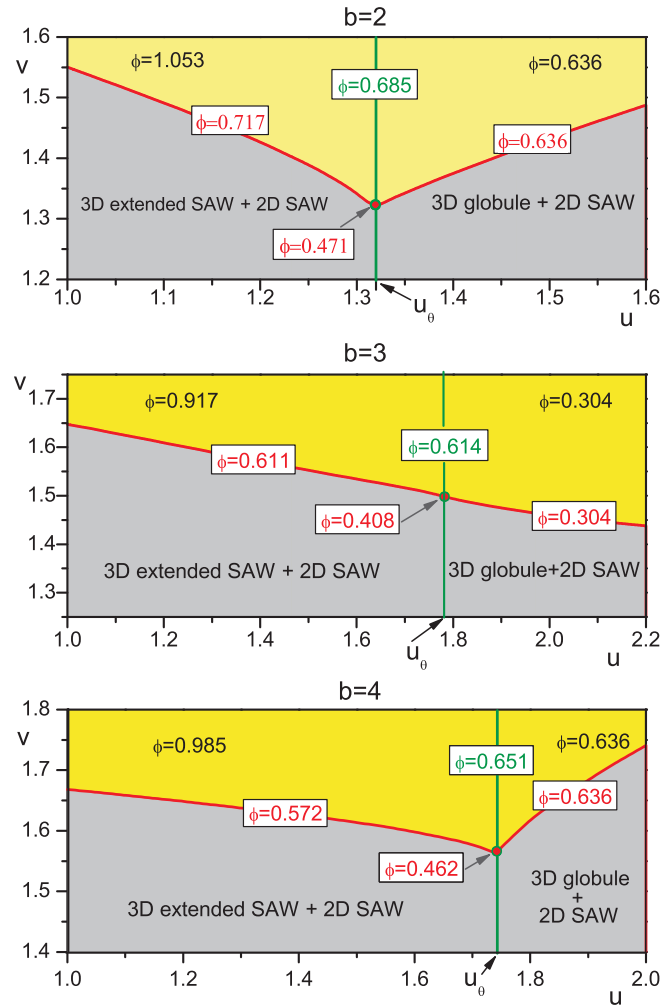} \end{center}
\caption{Phase diagrams obtained for the model of two avoiding
SAWs on 3D SG fractals with $b=2,3$ and 4. The solid vertical line
$u=u_\theta$ divides the $u-v$ plane in two areas, corresponding
to the phases in which the 3D SAW is either extended
($u<u_\theta$) or collapsed ($u>u_\theta$). Each of these two
areas is additionally partitioned by the critical line $v=v_c(u)$.
For $v<v_c(u)$ the two polymers are segregated one from another,
and the system exists either as ``3D extended SAW + 2D SAW" for
$u<u_\theta$, or ``3D globule + 2D SAW" for $u>u_\theta$, whereas
for $u=u_\theta$ precisely, $\theta$-chain coexists with 2D SAW.
For $v\geq v_c(u)$ the mean number $\langle M\rangle$ of contacts
between the two SAWs scales with the mean length $\langle
N_3\rangle$ of the 3D SAW as $\langle M\rangle\sim \langle
N_3\rangle^{\phi}$. Depending on the value of $u$, critical
exponent $\phi$ has different values, which are presented within
the corresponding areas.} \label{fig:FDAvoiding}
\end{figure}

To establish the exact  forms of RG equations, for  each fractal,
one needs to find the coefficients $a$, $b$, $c$, and $d$, that
appear in (\ref{eq:RGA})--(\ref{eq:RGA4}). Using the computer
facilities, by direct enumeration and classification of all
possible SAW configurations on the first stage of fractal
construction, it is feasible to find these coefficients for
fractals labelled by $b=2,3$ and 4 (see \ref{app:ASAWsRG}).
Precise numerical analysis of the obtained RG equations (for
$b=2,3$, and 4)  reveals that two-polymer system can reside in
several phases, depending on the values of the interaction
parameters $u$ and $v$. In particular, for each value of $u$,
there is a critical value $v=v_c(u)$, such that for $v<v_c(u)$ the
two chains exist almost independently in the solution. This is
indicated by the fact that $(A^*,B^*)$, and $C^*$ retain their
fixed values that correspond to the solitary chain on 3D SG, and
2D SG, respectively (see table~\ref{tab:avoiding}), and confirmed
by calculating the mean number of contacts $\langle
M^{(r)}\rangle$ between the chains, which quickly approaches some
constant value as $r\to\infty$.
{\Table{\label{tab:avoiding} Coordinates of transition fixed
points $(A^*, B^*, C^*, D^*)$, obtained via renormalization group
approach, for ASAWs model on 3D SG fractals labelled by $b=2,3$
and 4. Also, we give the corresponding relevant eigenvalues
$\lambda_{\nu_3}$ and $\lambda_{D}$, together with the pertaining
values of contact critical exponents $\phi$. For all values of $b$
studied, when $v<v_c(u)$, eigenvalue $\lambda_{D}$ is not
relevant, and the mean number of contacts between the chains is
finite (thus the values of $\lambda_{D}$ and $\phi$ are not given
for these cases).}
 \begin{tabular}{lclllllll}
 \br
 $b$ & $v$&$A^*$&$B^*$&$C^*$   & $D^*$
 &$\lambda_{\nu_3}$ & $\lambda_{D}$&$\phi$\\
\br
\multicolumn{8}{c}{extended 3D chain $(u<u_\theta)$}\smallskip\\
\mr
&$v<v_c(u)$&0.4294&0.0499&0.6180&0.1165&2.7965 &$<1$ &--\\
2 & $v=v_c(u)$& 0.4294&0.0499&0.6180&2.3303 &2.7965 &2.0904&0.7170\\
 & $v>v_c(u)$ &0.4294&0.0499&0&3.0887 &2.7965 &2.9537&1.0532\smallskip\\
\mr
&$v<v_c(u)$& 0.3420&0.0239&0.5511&0.0779&5.3620 &$<1$ &--\\
3 &$v=v_c(u)$&  0.3420&0.0239&0.5511&1.5388 &5.3620 & 2.7879&0.6105\\
 &$v>v_c(u)$ &  0.3420&0.0239&0&2.8591 &5.3620 & 4.6651&0.9171\smallskip\\
\mr
&$v<v_c(u)$&0.2899&0.0122&0.5063& 0.0580  &8.6911 &$<1$ &--\\
4 &$v=v_c(u)$&  0.2899&0.0122&0.5063&1.2051 &8.6911 &3.4427&0.5717\\
 &$v>v_c(u)$ &  0.2899&0.0122&0&2.0837 &8.6911 &8.4170&0.9852\\
 \br
 \multicolumn{8}{c}  {3D $\theta$--chain $(u=u_\theta)$} \\
 \mr
&$v<v_c(u_\theta)$&1/3&1/3&0.6180& 0.0613&100/27  &$<1$ &--\\
2& $v=v_c(u_\theta)$& 1/3&1/3&0.6180&0.6180 &100/27  &1.8526&0.4709\\
&$v>v_c(u_\theta)$&  1/3&1/3&0&0.8229 &100/27  &2.4514&0.6848\smallskip\\
\mr
&$v<v_c(u_\theta)$&0.2071&0.4307&0.5511&0.0211 &8.7231 &$<1$ &--\\
3&$v=v_c(u_\theta)$&  0.2071&0.4307&0.5511&0.5773 &8.7231 & 2.4203&0.4081\\
&$v>v_c(u_\theta)$&  0.2071&0.4307&0&1.2781 &8.7231 & 3.7800&0.6139\smallskip\\
\mr
&$v<v_c(u_\theta)$&0.1918&0.3393&0.5063& 0.0180 &15.424 &$<1$ &--\\
 4 &$v=v_c(u_\theta)$&  0.1918&0.3393&0.5063&0.5758 &15.424 &3.5367&0.4617\\
   & $v>v_c(u_\theta)$& 0.1918&0.3393&0&0.7984 &15.424 &5.9331&0.6508\\
  \br
 \multicolumn{8}{c}  {3D globule $(u>u_\theta)$}\smallskip\\
 \mr
&$v<v_c(u)$& 0&$22^{-1/3}$&0.6180& 0 &4&$<1$ &--\\
2&$v=v_c(u)$&  0&$22^{-1/3}$&0.6180&0.7637 &4   &2.4163&0.6364\\
&$v>v_c(u)$&  0&$22^{-1/3}$&0&0.7637 &4   &2.4163&0.6364\smallskip\\
\mr
&$v<v_c(u)$&0&$\infty$ &0.5511& 0 &9.772 &$<1$ &--\\
3&$v=v_c(u)$&  0&$\infty$ &0.5511&$\infty$ &9.772 & 2& 0.3041\\
&$v>v_c(u)$&  0&$\infty$ &0&$\infty$ &9.772 & 2& 0.3041\smallskip \\
\mr
&$v<v_c(u)$&0&$22^{-1/3}$&0.5063& 0&16 &$<1$ &--\\
4 &$v=v_c(u)$  &  0&$22^{-1/3}$&0.5063&0.7637 &16 &5.8387&0.6364 \\
 &$v>v_c(u)$&  0&$22^{-1/3}$&0&0.7637 &16 &5.8387&0.6364 \\
 \br
    \end{tabular}
\endTable}
For $v=v_c(u)$ fixed values $(A^*,B^*)$ and $C^*$ remain the same
as for $v<v_c(u)$, but $D^*$ becomes larger, and $\langle
M^{(r)}\rangle$ increases with $r$, obeying the scaling relation
(\ref{asawfi}), with eigenvalue $\lambda_D$ being larger than one.
Although there are large number of contacts between them, both
chains also have large parts that are not interconnected. Even for
$v>v_c(u)$ fixed value $(A^*,B^*)$ does not change, but then $C^*$
becomes equal to zero, meaning that the whole $P_2$ chain is
covered with the $P_3$ chain (which still has a lot of monomers in
the bulk, far from the boundary in which $P_2$ lies). The regions
$v>v_c(u)$ and  $v<v_c(u)$ of the phase plane $u-v$, as well as
the critical line $v_c(u)$, are additionally partitioned by the
vertical line $u=u_\theta$, so that each part obtained in such a
way is characterized by different fixed point $(A^*,B^*,C^*,D^*)$,
corresponding to different phase. Coordinates of all fixed points
are given in table~\ref{tab:avoiding}, whereas in
figure~\ref{fig:FDAvoiding} one can see obtained phase diagrams
for $b=2$, 3, and 4 SG fractals.

Shape of the line $v_c(u)$, as well as values of the exponent
$\phi$, show that the interplay between the intra- and inter-chain
interactions in the system under study is quite subtle. In the
$b=3$ case, $v_c(u)$ decreases monotonically with $u$, meaning
that stronger monomer-monomer attraction within the $P_3$ chain
eases its attaching to the $P_2$ chain. However, for $b=2$ and 4
fractals this is correct only for values of $u$ up to $u_\theta$,
where $v_c(u)$ has its minimum. For larger values of $u$, function
$v_c(u)$ monotonically increases with $u$, {\em i.e.} for
$u>u_\theta$ intra-chain prevails inter-chain interaction and
hinders attaching. Different behavior of the system on fractals
with $b=2,4$ and $b=3$ is due to the peculiar topology of these
lattices. Namely, although for $u>u_\theta$ polymer is in globular
phase, compactness of that structure is not always the same. Only
on $b=2$ SG the globule is completely compact, {\it i.e.} its
fractal dimension $d_f^G$ is equal to the fractal dimension
$d_f^{3D}$ of the lattice \cite{DharVannimenus}. In the $b=3$ and
4 cases  $d_f^G<d_f^{3D}$, but this quasi-compactness is much more
pronounced in the $b=3$ case \cite{Knezevic,EZM}, which brings
about different behavior of the system on the $b=3$ SG fractal.
Concerning the exponent $\phi$, which can be taken as a measure of
interconnection between the two chains, one can notice that it has
different values on the critical line $v_c(u)$ and in the region
$v>v_c(u)$, and in addition depends on intra-chain interaction
parameter $u$ (see table~\ref{tab:avoiding} and
figure~\ref{fig:FDAvoiding}). For each of the three studied
fractals, in the range $v>v_c(u)$ the inequality
$\phi(u<u_\theta)>\phi(u=u_\theta)>\phi(u>u_\theta)$ is satisfied.
Such inequality could have been expected, since it means that when
$P_3$ chain completely covers $P_2$ chain, the number of contacts
between them is smaller if structure of the $P_3$ chain is more
compact. On the line $v_c(u)$, however, chain $P_2$ is only
partially covered with $P_3$, so that some similar conclusion is
not plausible, which is indeed in accord with the calculated
values of $\phi$. Besides, it is interesting that for $b=2$ and 4,
the smallest value of $\phi$ is obtained for $u=u_\theta$, which
is not the case for $b=3$ fractal.

 Finally, one should note that in the case $b=3$, for the globular
 state of a solitary 3D chain ($u>u_\theta$), the coordinates of
 the corresponding fixed point are $A_G=0$ and $B_G=\infty$.
 Furthermore, a numerical analysis of  function $D^{(r)}$, in the
 range $v\ge v_c(u)$ reveals that $D^*=\infty$. Nevertheless, the
 relation $\langle M^{(r)}\rangle\sim\lambda_D^r$ and formula
 (\ref{eq:skaliranje}) are applicable, but with different
 meaning of $\lambda_D$. For the globule state of $b=3$  fractal,
 it was demonstrated \cite{Knezevic} that equations (\ref{eq:Ab3})
 and  (\ref{eq:Bb3}) in the vicinity of the corresponding fixed
 point $(0,\infty)$ have the following approximate form
\begin{equation}\label{knez1}
A^{(r+1)}= 320 (A^{(r)})^3 (B^{(r)})^6\, , \quad B^{(r+1)}= 4308 (A^{(r)})^2 (B^{(r)})^8\>,
\end{equation}
from which it follows $\lambda_{\nu_3}=\frac{\sqrt{73}+11}2=9.772$
and $\nu_3^G=\ln 3/\ln 9.772=0.4819$. Besides, for $v\ge v_c(u)$,
the inequality $D^{(r)}\ll B^{(r)}$ is valid, so that RG equation
(\ref{eq:A4b3}) obtains the approximate form
\begin{equation}\label{jedn2}
 D^{(r+1)}=320 A^{(r)} (B^{(r)})^6 C^{(r)} (D^{(r)})^2\, .
\end{equation}
Then, from equations (\ref{eq:srednjeMASAWs}) and (\ref{dda}), follows
$
\langle M^{(r+1)}\rangle=2\frac v{D^{(r)}}\frac{\partial D^{(r)}}{\partial v}=2\langle M^{(r)}\rangle
$,
implying that
$
\langle M^{(r)}\rangle\sim \lambda_D^r$ (for large $r$), with $\lambda_D=2$.
 Finally,  from (\ref{eq:skaliranje}), one obtains
 $\phi={\ln2}/{\ln9.772}=0.3041$.

\section{The model of crossing walks}
\label{CSAWs}

In order to describe the physical situation when closer contact
between the two polymers is possible, in this section we analyze
the CSAWs model in which chains $P_2$ and $P_3$ can cross each
other \cite{ZivicJSTAT}. If we assume that chains interact only at
the crossing sites, and, similarly as in the ASAWs case, introduce
the weight factor $w=e^{-\epsilon_c/k_BT}$, where
$\epsilon_c\leq0$ is the energy of two monomers in contact, it
turns out that the two chains cannot exist independently, even for
extremely weak attraction ($|\epsilon_c|\ll k_BT$). Therefore, we
define additional weight factor $t=e^{-\epsilon_t/k_BT}$, where
$\epsilon_t>0$ is the energy associated with two sites, visited by
different SAWs, and both neighbouring a crossing site (see figure
\ref{fig:interakcije}(b)), so that unbinding transition can occur.
To describe exactly all possible configurations of the two-chain
polymer system, within this  model we need to introduce nine
restricted partition functions: $A^{(r)}$, $B^{(r)}$, $C^{(r)}$,
$A_1^{(r)}$, $A_2^{(r)}$, $A_3^{(r)}$, $A_4^{(r)}$, $B_1^{(r)}$,
and $B_2^{(r)}$.
\begin{figure}
\hskip4cm
\includegraphics[scale=0.4]{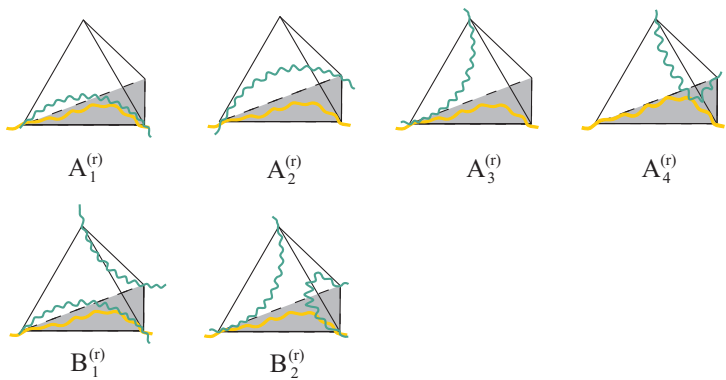}
\caption{The six
restricted generating functions used in the description of all
possible  inter-chain configurations for the CSAWs model of the
two-polymer system, within the $r$-th stage of 3D SG fractal
structure. The 3D  chain is  depicted by green line, while the 2D
surface-adhered
chain  is depicted by  yellow line.}
\label{figure4}
\end{figure}
Functions $A^{(r)}$, $B^{(r)}$ and $C^{(r)}$, which correspond to one-polymer configurations are the same as in the ASAWs model (see figure \ref{fig:RGparametri}, and RG relations (\ref{eq:RGA}) and (\ref{eq:RGB})), whereas the remaining six functions, which describe the inter-chain configurations,  are depicted in figure~\ref{figure4}, and they are defined as
\begin{eqnarray}
 A_i^{(r)}&=& \sum_{N_2,N_3,L,M,K}{\mathcal A}_i^{(r)}(N_2,N_3,L,M,K) x_2^{N_2}x_3^{N_3}u^L w^M t^K\, , \quad i=1,2,3,4\, ,\nonumber\\
 B_i^{(r)}&=&\sum_{N_2,N_3,L,M,K}{\mathcal B}_i^{(r)}(N_2,N_3,L,M,K) x_2^{N_2}x_3^{N_3}u^L w^M t^K\, , \quad i=1,2\, ,\nonumber
\end{eqnarray}
where ${\mathcal A}_i^{(r)}$ and ${\mathcal B}_i^{(r)}$ are the numbers of particular two-polymer configurations on the $r$-th fractal structure. For instance, ${\mathcal A}_4^{(r)}(N_2,N_3,L,M,K)$ is the number of configurations in which the $N_3$-step $P_3$ chain (with $L$ intra-chain contacts) and $N_2$-step $P_2$ chain (with different entering end exiting vertices from $P_3$ chain) cross $M$ times and have $K$ pairs of sites belonging to different chains and neighboring the crossing sites. Functions $A_i^{(r)}$ and $B_i^{(r)}$ satisfy the following recursion relations
\begin{eqnarray}
   A'_i&=&  \sum_{\cal{N}}
     a_i({\cal{N}})\,
 A^{N_A}B^{N_B}C^{N_C}
\prod_{j=1}^{4} A_{j}^{N_{A_j}}
\prod_{k=1}^2B_{k}^{N_{B_k}}\,,\quad i=1,2,3,4\>,
\label{eq:RGAi}\\
 B'_i&=&  \sum_{\cal{N}} b_i({\cal{N}})\,
 A^{N_A}B^{N_B}C^{N_C}
\prod_{j=1}^{4} A_{j}^{N_{A_j}}\prod_{k=1}^2
B_{k}^{N_{B_k}}\,,\quad i=1,2\>, \label{eq:RGBi}
\end{eqnarray}
where $\cal{N}$ denotes the set of numbers
${\cal{N}}=\{N_{A},N_{B},N_{C},N_{A_1},N_{A_2},N_{A_3,}N_{A_4},
N_{B_1},N_{B_2}\}$, and where we have used the prime symbol as a
superscript for $(r+1)$-th restricted partition functions and no
indices for the $r$-th order partition functions. The above set of
relations (\ref{eq:RGAi})--(\ref{eq:RGBi}), together with the
previously introduced relations (\ref{eq:RGA})--(\ref{eq:RGC}) for
the functions $A$, $B$, and $C$,  can be considered as the system
of RG equations for the problem under study, with the  initial
conditions
\begin{eqnarray}
 &&A^{(0)}=x_3\,,\quad B^{(0)}=x_3^2u^4\,,\quad C^{(0)}=x_2\,,\nonumber\\
 && A_1^{(0)}=x_3x_2w^2\,,\quad
A_2^{(0)}=A_3^{(0)}=x_3x_2wt\,,\quad
A_4^{(0)}=x_3x_2\,, \label{pocuslovi}\\
&&  B_1^{(0)}=B_2^{(0)}=x_3^2x_2w^2u^4\,,\nonumber
\end{eqnarray} corresponding to the unit tetrahedron.
Because the number of all possible configurations is  extremely
large,  we have been able to find explicit form of the RG
equations (\ref{eq:RGAi})--(\ref{eq:RGBi}) only for $b=2$ and
$b=3$ SG fractals (see \ref{app:CSAWsRG}). For both cases
numerical analysis shows that, for each considered value of $t$,
there is a critical line $w_c(u,t)$ dividing the $u-w$ plane into
regions where the two polymers are either segregated
($w<w_c(u,t)$) or entangled ($w\geq w_c(u,t)$). Depending on the
value of the intra-chain interaction parameter $u$, the area
$w\leq w_c(u,t)$ is further partitioned into smaller regions,
corresponding to various phases of the system (see figure~
\ref{fig:fdCSAWs}).
\begin{figure}
\hskip3cm
\includegraphics[scale=1.1]{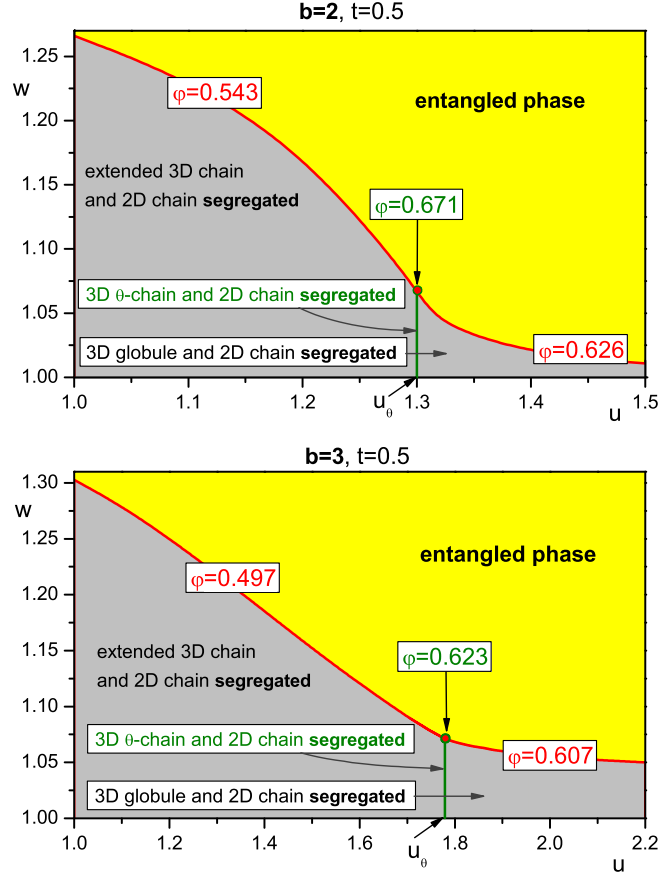}
\caption{Phase diagrams in the space of interaction parameters for
CSAWs model in the case of  $b=2$ and  $b=3$ SG fractal, for $t=0.5$.
 In both cases the critical line $w=w_c(u,t)$ separates the $u-w$ plane into the area $w>w_c(u,t)$ of entangled phase and area $w<w_c(u,t)$, in which the two chains are segregated. The latter area is divided by vertical line $u=u_\theta$ into regions, corresponding to three segregated phases: (i) 2D chain (always extended) and extended 3D chain ($u<u_\theta$), (ii) 2D chain and 3D $\theta$-chain  ($u=u_\theta$), and (iii) 2D chain and 3D globule ($u>u_\theta$).
One should observe that there appears  the multi-critical point
(full red circle) at the crossing of the $\theta$--line and the
critical line $w=w_c(u,t)$. For other values of $t$ ($0<t<1$),
the critical line $w_c(u,t)$ also monotonically decreases, for
both $b=2$ and $b=3$ fractals.}
 \label{fig:fdCSAWs}
\end{figure}
To each of these area different fixed point of the general type
\begin{equation}\label{fpgen}
    (A^*,B^*,C^*,A_1^*,A_2^*,A_3^*,A_4^*,B_1^*,B_2^*)\>,
\end{equation}
pertains. We describe general features of the fixed points and the
corresponding phases in the three following subsections.

\subsection{Weak self-attraction of the 3D chain}

For each value of $0<t<1$, and small values of the interaction parameter $1\leq u<u_\theta$, there is some critical value $w=w_c(u,t)$  such that
\begin{itemize}
\item For $w<w_c(u,t)$ the fixed point of the form
\begin{equation}\label{fp1}
(A_E,B_E,C^*,0,0,0,A_{4}^*,0,0)\>,
\end{equation}
is reached. This point corresponds to the phase in which 2D chain and
extended 3D chain are segregated, since as it is
approached, after some number $r\gg1$ of RG iterations, the
average number of contacts between the two
chains, quickly becomes constant. Values of $A_E$ and $B_E$ are fixed values of the RG parameters for the solitary extended chain on 3D SG fractal, and they are presented in table~\ref{tab:CSAWs}, together with the values of $C^*$, corresponding to 2D chain, which can exist only in extended state. RG fixed point value $A_4^*$ is equal to $0.1165$ and $0.0779$, for $b=2$ and 3 respectively, and they coincide with the values of $D^*$ for $v<v_c(u<u_\theta)$ case in the ASAWs model (see table~\ref{tab:avoiding}).
\item For $w=w_c(u,t)$ one obtains the symmetrical fixed point
\begin{equation}\label{fp2}
(A_E,B_E,C^*,A_EC^*,A_EC^*,A_EC^*,A_EC^*,B_EC^*,B_EC^*)\>,
\end{equation}
which appears to be a tricritical fixed point. As one approaches this fixed point,  the average number of contacts ${\langle M^{(r)}\rangle}$ becomes infinitely large (although large parts of $P_2$ and $P_3$ are not in contact), and it turns out that it scales with the average length  ${\langle {N_3}^{(r)}\rangle}$ of the 3D chain, according to the power law
\begin{equation}\label{ficsaw}
{\langle M^{(r)}\rangle}\sim \langle
{N_3}^{(r)}\rangle^{\varphi}\>.
\end{equation}
To calculate the contact critical exponent $\varphi$, within the CSAWs model, we find the average number of contacts between two chains at the $r$th stage of fractal construction, through  the formula
\begin{eqnarray}
\langle M^{(r)}\rangle&=& {\sum_{N_2,N_3,L,M,K}M\left(\sum_{i=1}^4{\mathcal A}_i^{(r)}+{\sum_{j=1}^2{\mathcal B}_i^{(r)}}\right) x_2^{N_2}x_3^{N_3}u^L w^M t^K\over \sum_{i=1}^4 A_i^{(r)}+\sum_{j=1}^2 B_j^{(r)}}\nonumber\\
 &=& {w\over \sum_{i=1}^4 A_i^{(r)}+\sum_{j=1}^2 B_j^{(r)}}
 \left(\sum_{i=1}^{4}\frac{\partial A_i^{(r)}}{\partial w}+
 \sum_{j=1}^{2}\frac{\partial B_j^{(r)}}{\partial w}\right)\nonumber\\
 &=&
 {w\over \sum_{i=1}^6 X_i^{(r)}}
 \sum_{i=1}^{6}\frac{\partial X_i^{(r)}}{\partial w}
 \>,
\end{eqnarray}
where $X_i=A_i$ ($i=1,2,3,4$), $X_5=B_1$, and $X_6=B_2$.
Since
\begin{equation}\label{nov1}
\frac{\partial X_i^{(r+1)}}{\partial w}=\sum_{j=1}^6 \frac{\partial X_i^{(r+1)}}{\partial X_j^{(r)}}\frac{\partial X_j^{(r)}}{\partial w}\>,\quad i=1,\ldots,6\>,
\end{equation}
one expects, for large $r$, that $\frac{\partial X_i^{(r)}}{\partial w}$ behaves as $\lambda_\varphi^r$, where $\lambda_\varphi$ is the largest relevant solution of the  eigenvalue equation
\begin{equation}\label{csawlamdafi}
  \mbox{det}\left| \left({\partial X^{(r+1)}_i\over \partial X^{(r)}_j}
    \right)^{*}-
    \lambda_\varphi\,\delta_{ij} \right|=0\>,
\end{equation}
where the asterisk means that the derivatives should be taken at
the tricritical fixed point.
  From here follows
$\langle M^{(r)}\rangle\sim \lambda_\varphi^r$, which together  with   $\langle N_3^{(r)}\rangle\sim \lambda_{\nu_3}^r$ (where $\lambda_{\nu_3}$, as before, is the largest eigenvalue of the
linearized RG equations for the bulk parameters $A$ and $B$), and (\ref{ficsaw}),
gives
\begin{equation}\label{fie2}
\varphi=\frac{\ln\lambda_{\varphi}}
{\ln\lambda_{\nu_3}}\>.
\end{equation}

\item For larger values of the interaction parameter $w>w_c(u,t)$,
 the RG parameters flow towards the fixed point
\begin{equation}\label{fp5}
(0,0,0,A_1^*=C^*,0,0,0,0,0)\>,
\end{equation}
which describes the entangled phase of the two polymers, in which $P_3$ chain is completely attracted to $P_2$ chain.
\end{itemize}
\begin{table}{\caption{\label{tab:CSAWs}  The CSAWs model  fixed points
corresponding to the critical values $w=w_c(u,t)$  of the
attraction parameter between the 2D and 3D chains, when $0<t<1$,
for all possible states of the 3D polymer, together with the
values of the critical exponent $\varphi$.}}
\raggedleft
\footnotesize\rm
\begin{tabular}{lllllllllll}
 \br
 $b$ & $A^*$&$B^*$&$C^*$   & $A_1^*$ & $A_2^*$ &$A_3^*$&$A_4^*$
 &$B_1^*$ & $B_2^*$&$\varphi$\\
\br
\multicolumn{11}{c}{extended 3D chain $(u<u_\theta)$}\\
\mr
2 &  0.4294&0.0499&0.6180&0.2654 &0.2654&0.2654&0.2654&0.0308
&0.0308&0.5428
    \\
    3 & 0.3420&0.0239&0.5511& 0.1884& 0.1884&
     0.1884& 0.1884&0.0131&0.0131&0.4973
    \\ \br
 \multicolumn{11}{c}
  {3D $\theta$--chain $(u=u_\theta)$} \\ \mr
    2& 1/3 & 1/3&0.6180& 0.0510&0&0&0.0613
    &0.2365&0.2362&0.6714\\
     3& 0.2071 & 0.4307 &0.5511&0.0810&0.0310&0.0250&0.0270
    &0.3130&0.3150&0.6226\\
    \br
 \multicolumn{11}{c}
  {3D globule $(u>u_\theta)$}\\ \mr
    2 & 0 & 0.3569 & 0.6180&0&0&0&0&0.2206& 0.2206&0.6261\\
    3 & 0 & $\infty$ &0.5511& 0&0&0&0&$\infty$&$\infty$&0.6073\\ \hline\hline
    \end{tabular}
\end{table}

\subsection{Critical self-attraction of the 3D chain}

For $u=u_\theta$
the solitary 3D chain is in the state of  the $\theta$-chain, for which $(A^*,B^*)=(A_\theta,B_\theta)$, whereas the two-polymer system can be in the following phases:
\begin{itemize}
\item For $w<w_c(u_\theta,t)$ the corresponding fixed point is of the form
\begin{equation}\label{fpt1}
(A_\theta,B_\theta,C^*,0,0,0,A_4^*,0,0)\>.
\end{equation}
This is the case when the 3D $\theta$-chain is segregated from the 2D chain
chain. Fixed point values of $A_4^*$ are: 0.0613 for $b=2$, and 0.0211 for $b=3$ fractal, equal to $D^*$ for the corresponding cases $v<v_c(u_\theta)$ of the ASAWs model.
\item When $w=w_c(u_\theta,t)$, the RG
parameters tend to the fixed point
\begin{equation}\label{fpt2}
  (A_\theta,B_\theta,C^*,A_1^*,A_2^*,A_3^*,A_4^*,B_1^*,
 B_2^*)\>,
\end{equation}
which corresponds to the phase in which chains are not segregated
anymore, but they are not yet completely entangled. In contrast to
the $w=w_c(u<u_\theta,t)$ case, for which symmetrical fixed point
is obtained, values of $A_i$ $(i=1,2,3,4)$, as well as $B_1$ and
$B_2$, are not mutually equal ($A_i\neq A_\theta C^*$, $B_i\neq
B_\theta C^*$). The scaling relation $\langle M^{(r)}\rangle\sim
\langle {N_3}^{(r)}\rangle^{\varphi}$ is satisfied, with $\varphi$
given by (\ref{fie2}).

\item For $w>w_c(u_\theta,t)$  the RG parameters flow towards the entangled fixed
point (\ref{fp5}).
\end{itemize}

\subsection{Strong self-attraction of the 3D chain}

When self-attraction of the 3D polymer is strong
($u>u_\theta$), depending on the values of inter-chain
interaction parameters, the following phases are possible:
\begin{itemize}
\item For $w<w_c(u,t)$ the chains are segregated. Due to the large compactness
of the 3D chain, with $(A^*,B^*)=(0,B_G)$, none of the configurations
$A_1,A_2,A_3,A_4,B_1,B_2$ can be accomplished, and the
corresponding fixed point is
\begin{equation}\label{fpg1}
(0,B_G,C^*,0,0,0,0,0,0)\, .
\end{equation}
The chains are completely separated.
\item When attraction between the chains is critical,
$w=w_c(u,t)$,  the chains are partially entangled, and the fixed point
\begin{equation}\label{fpg2}
(0,B_G,C^*,0,0,0,0,B_G C^*,B_G C^*)\>,
\end{equation}
is attained. In this case the interaction between chains is
sufficiently strong to connect them, but not  strong enough to
destroy the compactness of the 3D globule, so that all $A_i^*=0$.
Again, the scaling relation $\langle M^{(r)}\rangle\sim \langle
{N_3}^{(r)}\rangle^{\varphi}$ is satisfied for both $b=2$ and 3,
with $\varphi$ given by (\ref{fie2}). However, while in the case
$b=2$ the coordinates of corresponding fixed point have definite
values
 (and $\lambda_\varphi$ can be directly calculated from linearizied RG equations  for $A_i$ and $B_i$),
 in the $b=3$ case   some fixed point coordinates
  diverge, and calculation of $\lambda_\varphi$ requires an additional effort. To be more specific, a numerical analysis of RG equations (\ref{eq:RGAi}) and (\ref{eq:RGBi}) reveals that $A_i^{(r)}\approx A^{(r)}C^*\to 0$, $B_i^{(r)}\approx B^{(r)}C^*\to \infty$. In this situation
  the appropriate eigenvalue $\lambda_\varphi$ can also be calculated, using the following transformation.
   If we write the relation (\ref{nov1})  in the form
\begin{equation}\label{nov2}
\fl{1\over X_i^{(r+1)}}\frac{\partial X_i^{(r+1)}}{\partial w}=\sum_{j=1}^6 \left({X_j^{(r)}\over X_i^{(r+1)}}\frac{\partial X_i^{(r+1)}}{\partial X_j^{(r)}}\right){1\over X_j^{(r)}}\frac{\partial X_j^{(r)}}{\partial w}\>,\quad i=1,\ldots,6\>,
\end{equation}
it can be shown that, when we keep only dominant terms in the RG
equations, and in the derivatives  $\frac{\partial
X_i^{(r+1)}}{\partial X_j^{(r)}}$, then, the  matrix elements
${\left(\frac{X_j^{(r)}}{X_i^{(r+1)}}\frac{\partial
X_i^{(r+1)}}{\partial X_j^{(r)}}\right)}^*$ of the new eigenvalue
problem are  either equal to zero or to some finite constants
(depending on $C^*$), from which we  find
$\lambda_\varphi=3.9919$, and therefrom
$\varphi={\ln\lambda_\varphi}/{\ln\lambda_{\nu_3}}=0.6073$.

\item Strong inter-chain attraction $w>w_c(u,t)$ destroys
the globule and completely attaches the 3D chain to the 2D chain.
This entangled phase is again characterized by the fixed point
(\ref{fp5}).
\end{itemize}

\begin{figure}
\hskip2cm
\includegraphics[scale=1.1]{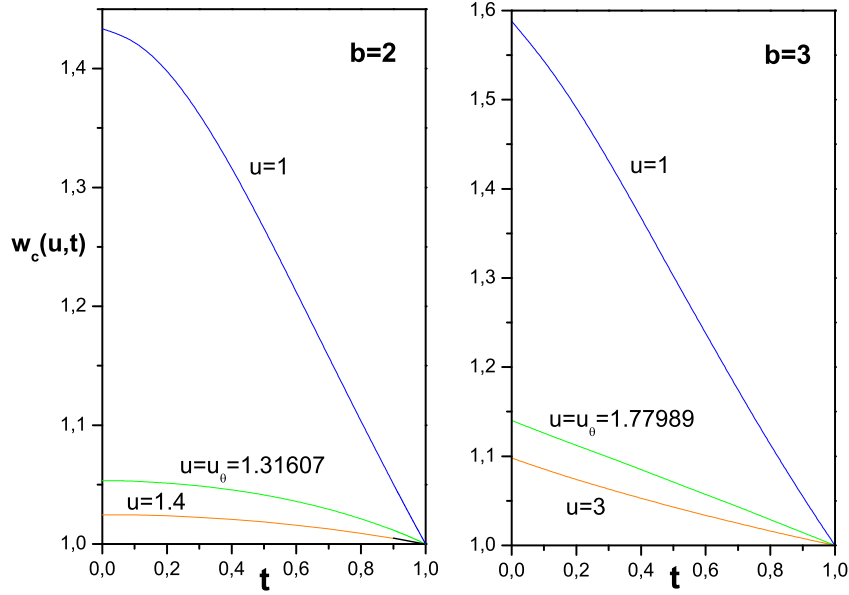}
\caption{Critical value of the inter-chain interaction
parameter $w_c(u,t)$, depicted as a function of $t$, for  three
different values of intra-chain interaction parameter $u$, in the
cases of $b=2$ and $b=3$ 3D SG fractals. }
 \label{fig:wcODt}
\end{figure}

In table~\ref{tab:CSAWs} we have presented the numerical results
for the crossover fixed points and the corresponding values of the
contact exponent $\varphi$, obtained for the unbinding transitions
from entangled two-polymer phase to segregated phases of 2D and 3D
chains on the  $b=2$ and $b=3$ 3D SG fractals. These values are
correct for all studied cases of $t$ in the interval $(0,1)$.
Varying the parameter $t$ in this interval changes only the
particular values of $w_c(u,t)$, but, for both $b=2$ and $b=3$,
the function $w_c(u,t)$ for fixed $t$ is monotonically decreasing
function (see figure~\ref{fig:fdCSAWs}). Dependence of $w_c(u,t)$
on $t$, when $u$ is fixed, is presented in figure~\ref{fig:wcODt},
for several values of $u$.
As one can see, the limiting values $t=0$  and $t=1$  are also
included in this figure. However, in these cases different fixed
points, from those obtained for $0<t<1$, can be reached, which  is
expounded in the following paragraphs.

First, we analyze the value $t=0$, which represents  the limiting
case, within the CSAWs model, when the energy $\varepsilon_t$
(corresponding to the repelling of two different chain monomers,
placed at  sites which are nearest neighbours to a crossing site)
is infinitely large. Starting with the initial values
(\ref{pocuslovi}), it can be shown that, in the case of the $b=3$
fractal, the same fixed points of the RG equations (\ref{eq:RGAi})
and (\ref{eq:RGBi}), as for $0<t<1$ are reached. However, for the
$b=2$ fractal, it can be seen, from the explicit form of the RG
equations (\ref{eq:b2jednacinaA1})--(\ref{eq:b2jednacinaB2}), that
$t=0$ leads to $A_2^{(r)}=A_3^{(r)}=0$, for every $r$, $x_2$,
$x_3$, $u$ and $w$. This is due to the topology of this fractal,
and a consequence is that the fixed point
$(A_E,B_E,C^*,C^*,0,0,A_4^*,A_4^*,0)$ corresponds to the critical
values $w=w_c(1\leq u< u_\theta,t=0)$. The coordinates of this
fixed point $A_E=A^*$, $B_E=B^*$ and $C^*$ are given in the part
``extended 3D chain ($u<u_\theta$)" of the table~\ref{tab:CSAWs},
while $A_4^*=0.1164$, and the concomitant  critical exponent is
$\varphi=0.8439$. The remaining fixed points are the same as for
$0<t<1$.

The second limiting value ($t=1$) corresponds to the case
$\varepsilon_t=0$ (when  there is no repelling interaction). In
this case, for all $u$,  the critical value of the interaction
parameter $w$ is equal to $w_c(u,t=1)=1$. This means that the
chains can not be segregated, even for extremely small attraction
between them. For both fractals $b=2$ and $b=3$, the fixed points
that pertain to the critical value $w_c$, for $u\neq u_\theta$,
are the same as for $0<t<1$. For $u=u_\theta$ the symmetrical
fixed point is reached, $(A_\theta,B_\theta,C^*,A_\theta
C^*,A_\theta C^*,A_\theta C^*,A_\theta C^*,B_\theta C^*,B_\theta
C^*)$, in contrast to the case $0<t<1$. Values of $A_\theta=A^*$,
$B_\theta=B^*$ and $C^*$ can be found in the middle part of the
table~\ref{tab:CSAWs}, while the values of the contact critical
exponents are  $\varphi(b=2)=0.6102$, and $\varphi(b=3)=0.5907$.

Finally, one should mention that recently, using the Monte Carlo
renormalization group (MCRG)  method, the contact exponent
$\varphi$  was calculated for $b$ up to 40, for the case when the
intra-chain interactions within the 3D chain are negligible, $u\to
0$  \cite{ZivicJSTAT}. Comparing the reported MCRG data
$\varphi^{MC}(b=2)=0.5440\pm0.0056$ and
$\varphi^{MC}(b=3)=0.4969\pm0.0024$ with our exact findings
$\varphi(b=2)=0.5428$ and $\varphi(b=3)=0.4973$, we can see that
MCRG data are in excellent agreement with our exact findings. In
\cite{ZivicJSTAT} it was also demonstrated that $\varphi^{MC}$, as
a function of the scaling parameter $b$, continues decreasing with
increasing $b$, and, it seems that in the fractal-to-Euclidean
crossover region (\textit{i.e.} in the limit $b\to\infty$) it goes
to the proposed zero Euclidean value. The inequality
$\varphi(b=2)>\varphi(b=3)$ is satisfied not only for weak
intra-chain interactions ($u<u_\theta$), but also for $u\geq
u_\theta$, as can be seen in table~\ref{tab:CSAWs}. Unfortunately,
in the range $u\geq u_\theta$, the MCRG calculation of $\varphi$
is not feasible,   so that  prediction for the large $b$ behavior
of $\varphi(u\geq u_\theta)$, only on the bases of our exact data,
is not possible.

\section{Summary and conclusion}
\label{sumiranje}

In this paper we have studied a system of two interacting
chemically different  polymer chains in a  poor solvent. Such a
situation can be modelled by two avoiding self-avoiding walks
(ASAWs), as well as by two crossing self-avoiding walks (CSAWs).
We assume that polymers are situated in  fractal containers
modelled by members of 3D SG fractal family, which are labelled by
an integer $b$ ($2\le b<\infty$). We adopt that the first polymer
($P_3$) is a floating chain in the bulk of 3D SG fractal, while
the second ($P_2$) is stuck to one of the four boundaries of the
3D SG fractal, which appears to be a 2D SG fractal. To take into
account the  intra-chain interaction of $P_3$ polymer we have
introduced the interaction parameter $u={\mathrm{
e}}^{-\varepsilon_{u}/k_BT}$, where $\varepsilon_{u}<0$ is the
energy corresponding to interaction between two nonconsecutive
neighboring monomers within the chain. In the case of ASAWs model,
the two SAW paths cannot intersect each other, and  we assume that
two polymers interact when they approach  a distance  equal to a
lattice constant. We associate the weight factor
$v=e^{-\epsilon_v/k_BT}$ with each such contact, where
$\epsilon_v<0$ is the appropriate energy of inter-chain
interaction. On the other hand, in the case of CSAWs model, in
order to describe the inter-chain interactions of $P_3$ and $P_2$,
we have introduced the parameters
$w={\mathrm{e}}^{-\varepsilon_{c}/k_BT}$ and
$t={\mathrm{e}}^{-\varepsilon_t/k_BT}$, where $\epsilon_c<0$ is
the energy corresponding to each crossing of SAWs, while
$\epsilon_t>0$ is the energy associated with a pair of sites,
visited by different SAWs, which are nearest neighbors to a
crossing site.

To obtain the phase diagrams and the contact critical exponents
between the two polymers, we have applied an exact RG method for
the 3D SG fractals labelled by $b=2,3$ and 4, in the case of ASAWs
model, and for fractals $b=2$ and 3, in the case of CSAWs model.
In both models, for various values of intra-chain interaction
parameter $u$, a solitary 3D floating polymer chain can be found
in one of the three possible phases (extended, $\theta$-phase, or
globule phase), whereas a solitary 2D chain is always extended.
Depending on the values of the inter-chain interaction parameters
($v$ in the case of ASAWs model, and $w$ and $t$ in the case of
CSAWs model), the system can be either in the segregated phase,
when the chains can be considered as almost independent, or in
phases in which the number of contacts between the chains is
comparable with their length (entangled phases). For both models,
there is a critical line in the plane of the interaction
parameters ($v_c(u)$ for ASAWs, and, $w_c(u,t)$, with fixed $t$,
for CSAWs model), which divides it into areas corresponding to
segregated and entangled phases. In the case of the ASAWs model,
for $v\geq v_c(u)$, the average number $\langle M\rangle$ of
contacts between the two polymers scales with the average length
$\langle N_3\rangle$ of the 3D chain as $\langle M\rangle \sim
{\langle N_3\rangle}^\phi$. Different values of the contact
exponent $\phi$ correspond to $v=v_c(u)$ and $v>v_c(u)$, and, in
addition, $\phi$ also depends on the strength of the intra-chain
interaction parameter $u$. However, in all entangled phases large
parts of the 3D chain remain in the bulk, beyond the scope of the
inter-chain interaction, since the prohibition of crossings
between two chains hinders their complete entanglement, even for
extremely large values of $v$. On the contrary, in CSAWs model for
$w>w_c(u,t)$ the two chains are completely entangled, while they
only partially cover each other at the critical line $w=w_c(u,t)$,
where the scaling relation $\langle M\rangle \sim {\langle
N_3\rangle}^\varphi$ is satisfied, and where $\varphi$ takes
different values in the intra-chain interaction regions
$u<u_\theta$, $u=u_\theta$, and $u>u_\theta$.

In the end, we would like to point out that for ASAWs model, in
the space of interaction parameters, the  arrangement of possible
phases  is approximately the same, as in the case of the
surface-interacting polymer chain in a poor solvent in Euclidean
spaces \cite{r1,r2,r3}. On the other hand, the obtained phase
diagrams for CSAWs model, resemble the phase diagrams of the same
surface-interacting chain problem, in fractal containers
\cite{EZM}. This similarity is not surprising, since in both
models studied, one of the two interacting polymers is adhered to
one of four fractal surfaces, and its monomers appear as a part of
interacting surface (in the surface-interacting polymer problem).
Here we may conclude that, our findings should be useful in making
the corresponding 3D models of the system of several interacting
polymer chains in porous media. Besides, our results may serve
inspiring in advancing theories of mutually interacting polymers,
as well as for polymers interacting with boundary surfaces of
homogeneous 3D lattices, in which case so far (to the best of our
knowledge) an exact approach has not been yet made.

\ack
{This paper has been produced as  part of the work within the
project No.141020B funded by the Serbian Ministry of Science and
Protection of the Living Environment.}

\appendix

\section{Renormalization group equations for the ASAWs model \label{app:ASAWsRG}}

In this Appendix we give explicit RG equations for the model in
which two chains avoid each other, for the cases $b=2$, and $b=3$
of 3D SG fractals. Equations for ``bulk" parameters $A$ and $B$,
as well as for the ``surface" parameter $C$, were found in earlier
works, and we give them here only for the sake of completeness.

First, we give the RG equations for $b=2$ 3D SG fractal
\begin{eqnarray}
A'&=&A^2 + 2\,A^3 + 2\,A^4 + 4\,A^3\,B + 6\,A^2\,B^2\, , \label{eq:Ab2}\\
B'&=&A^4 + 4\,A^3\,B + 22\,B^4\, ,\label{eq:Bb2}\\
C'&=&C^2 + C^3\, ,\label{eq:Cb2}\\
D'&=&  2\,{ D}^3\,B + 6\,{ D}^2\,B^2 + 2\,A^2\, D\,C +
 A^2\,C^2 + A\, D\,C^2\, . \label{eq:Db2}
\end{eqnarray}
We note that first three equations were established for the first time in \cite{dhar78}.

Next, we present RG equations for the $b=3$ case
\begin{eqnarray}
\fl A'&=&A^3+6 A^4+16 A^5+34 A^6+76 A^7+112 A^8+112 A^9+ 64
A^{10}+ 8 A^4 B+ 36 A^5 B\nonumber\\ \fl
&+& 140 A^6 B+292 A^7 B +424 A^8 B+ 332 A^9 B+12 A^3
B^2+12 A^4
B^2+ 118 A^5 B^2\nonumber\\ \fl
&+&  380 A^6 B^2+ 806 A^7 B^2+664 A^8 B^2+72 A^4
B^3 +352 A^5 B^3+704 A^6 B^3+ 1728 A^7 B^3
 \nonumber\\ \fl
&+& 344 A^4 B^4+ 1568 A^5 B^4+848
A^6B^4+264 A^4 B^5+3192 A^5 B^5+  320 A^3 B^6\, , \label{eq:Ab3} \\
\fl B'&=&A^6+12 A^7+40 A^8+60 A^9+32 A^{10}+28 A^6 B + 88 A^7
B+224 A^8 B+160 A^9 B\nonumber\\
\fl &+& 40 A^6 B^2+496 A^7 B^2 +596 A^8 B^2 + 176 A^5
B^3 +768 A^6
B^3+ 1056 A^7 B^3+ 88 A^3 B^4\nonumber\\ \fl
&+&  264 A^5 B^4 + 2534 A^6 B^4+
1152 A^4 B^5+1888 A^5 B^5\nonumber\\ \fl &+& 5808 A^4 B^6+1936
A^3 B^7+ 4308 A^2 B^8 \, , \label{eq:Bb3}\\
\fl C'&=&C^3+3 C^4+C^5+2C^6\, , \label{eq:Cb3}
\\
\fl D'&=&2A^6 D^3 + 4 A^7 D^3 + 4 A^6 D^4 +
  2 A^5 D^5 + 28 A^6 D^3 B + 4 A^4 D^4 B +
  14 A^5 D^4 B  \nonumber
  \\ \fl&+& 4 A^4 D^5 B +
  8 A^4 D^3 B^2 + 56 A^5 D^3 B^2 +
  44 A^4 D^4 B^2 + 4 A^3 D^5 B^2  \nonumber\\
  \fl &+&
  144 A^4 D^3 B^3 + 36 A^3 D^4 B^3 +
  12 A^2 D^5 B^3 + 72 A^3 D^3 B^4 +
  132 A^2 D^4 B^4  \nonumber\\
  \fl&+& 264 A^2 D^3 B^5 +
  12 A^6 D^2 C + 18 A^7 D^2 C +
  2 A^4 D^3 C + 8 A^5 D^3 C + 8 A^6 D^3 C  \nonumber
  \\   \fl &+&    4 A^5 D^4 C + 2 A^4 D^5 C +
  16 A^4 D^2 B C + 48 A^5 D^2 B C +
  64 A^6 D^2 B C  \nonumber\\
  \fl &+& 4 A^2 D^3 B C +
  8 A^4 D^3 B C + 48 A^5 D^3 B C +
  4 A^3 D^4 B C + 8 A^4 D^4 B C \nonumber\\
  \fl &+&
  8 A^3 D^5 B C + 12 A^2 D^2 B^2 C +
  36 A^4 D^2 B^2 C + 162 A^5 D^2 B^2 C + 24 A^3 D^3 B^2 C\nonumber\\
  \fl &+& 28 A^4 D^3 B^2 C +
  44 A^3 D^4 B^2 C + 8 A^2 D^5 B^2 C + 96 A^3 D^2 B^3 C + 64 A^4 D^2 B^3 C \nonumber\\
  \fl &+&  152 A^3 D^3 B^3 C + 24 A^2 D^4 B^3 C +
  24 A D^5 B^3 C + 512 A^3 D^2 B^4 C \nonumber\\
  \fl &+&88 A D^4 B^4 C + 264 A^2 D^2 B^5 C +
  320 A D^2 B^6 C + 6 A^4 D C^2 + 16 A^5 D C^2 \nonumber\\
  \fl &+& 28 A^6 D C^2 + 12 A^7 D C^2 +
  12 A^5 D^2 C^2 + 18 A^6 D^2 C^2 + 2 A^3 D^3 C^2 \nonumber\\\fl&+&
  6 A^4 D^3 C^2 +
  8 A^5 D^3 C^2 + 4 A^4 D B C^2 + 64 A^5 D B C^2 + 56 A^6 D B C^2\nonumber\\
  \fl &+&
  12 A^3 D^2 B C^2 + 30 A^4 D^2 B C^2 +
  36 A^5 D^2 B C^2 + 2 A D^3 B C^2 +6 A^3 D^3 B C^2  \nonumber\\
  \fl &+&  24 A^4 D^3 B C^2+
  100 A^4 D B^2 C^2 + 88 A^5 D B^2 C^2 +
  6 A D^2 B^2 C^2  \nonumber\\
  \fl &+&  18 A^3 D^2 B^2 C^2 +
  56 A^4 D^2 B^2 C^2 + 4 A^2 D^3 B^2 C^2 +
  20 A^3 D^3 B^2 C^2  \nonumber\\
  \fl &+&  160 A^4 D B^3 C^2+
  36 A^2 D^2 B^3 C^2 + 32 A^2 D^3 B^3 C^2 +256 A^3 D B^4 C^2 \nonumber\\
  \fl &+& 132 A^2 D^2 B^4 C^2 +
  44 A D^3 B^4 C^2 + A^3 C^3 + 6 A^4 C^3 + 10 A^5 C^3 + 10 A^6 C^3  \nonumber\\
  \fl &+&  6 A^7 C^3 + 8 A^3 D C^3 + 10 A^4 D C^3 +
  14 A^5 D C^3 + 10 A^6 D C^3 \nonumber\\
  \fl &+&4 A^4 D^2 C^3 +
  6 A^5 D^2 C^3 + 2 A^3 D^3 C^3 +
  6 A^4 D^3 C^3 + 8 A^4 B C^3 + 16 A^5 B C^3\nonumber\\
  \fl &+& 20 A^6 B C^3 +
  28 A^4 D B C^3 + 36 A^5 D B C^3 +
  8 A^3 D^2 B C^3 + 12 A^4 D^2 B C^3 \nonumber\\
  \fl &+&
  4 A^2 D^3 B C^3 + 16 A^3 D^3 B C^3 + 12 A^3 B^2 C^3 +
  18 A^5 B^2 C^3 + 12 A^2 D B^2 C^3  \nonumber\\
  \fl &+& 44 A^4 D B^2 C^3 +
  12 A^3 D^2 B^2 C^3 + 20 A^2 D^3 B^2 C^3 +
  24 A^4 B^3 C^3\nonumber\\
  \fl &+& 48 A^3 D B^3 C^3  + 24 A^2 D^2 B^3 C^3 + 24 A D^3 B^3 C^3
   + 3 A^3 C^4 + 10 A^4 C^4  \nonumber\\
   \fl &+& 10 A^5 C^4 + 4 A^6 C^4 +A^2 D C^4
  + 2 A^3 D C^4 + 2 A^4 D C^4 +
  12 A^4 B C^4  \nonumber\\
  \fl &+&  8 A^5 B C^4 + 4 A^3 D B C^4 + 18 A^3 B^2 C^4 + 6 A^2 D B^2 C^4 + 2 A^2 D C^5\nonumber\\\fl&+&
   4 A^3 D C^5 +
  4 A^4 D C^5 + 8 A^3 D B C^5 + 12 A^2 D B^2 C^5 \> . \label{eq:A4b3}
\end{eqnarray}
Equations (\ref{eq:Ab3}) and (\ref{eq:Bb3}) were found in \cite{Knezevic}, and (\ref{eq:Cb3}) in \cite{EKM}.

For the $b=4$ case, equations are too cumbersome to be quoted
here, and, they  are available upon request to the authors.

\section{Renormalization group equations for the CSAWs model \label{app:CSAWsRG}}

It can be shown, via direct computer enumeration of the
corresponding paths within the generator of the $b=2$ 3D SG
fractal, that RG parameters $A_1, A_2, A_3, A_4, B_1$, and $B_2$
fulfil the following recursion relations
\begin{eqnarray}
\fl  A'_1&=&A_1^2 + A_1^3 + A{A_2^2} + 2A^2A_2A_3 + A{A_3^2} +
  2AA_1{A_3^2} + 2AA_2A_3B_1 + 2AA_2A_3B_2  \nonumber\\
  \fl &+&  4A^2A_1B_2 + 4A^2B_1B_2 +
  4AA_1B_1B_2 + 2A^2{B_2^2}+ 2AA_1{B_2^2} +
  {A_2^2}C + A{A_3^2}C\>, \label{eq:b2jednacinaA1}\\
\fl   A'_2&=&AA_1A_2 + A_2^3 + A^2A_1A_3 + AA_2A_3^2 +
  A^2A_3A_4 + AA_2A_4^2 + 2AA_2A_4B+ AA_3A_4C\nonumber\\
   \fl &+&  A^2A_3B_1+
   AA_1A_3B_1 + 2AA_2BB_1 +
  AA_2B_1^2 + A^2A_3B_2 +4AA_2BB_2 +A^2A_3C
 \nonumber\\
 \fl &+&   2AA_2B_1B_2 + 3AA_2B_2^2 + AA_2C +
   A_1A_2C + AA_1A_3B_2 +
  2AA_2A_4B_2   \, ,\\
 \fl  A'_3&=&A^2A_1A_2 + AA_1A_3 + AA_1^2\, A_3 +
  AA_2^2A_3 + A^2A_2A_4 + 2A_3^3B+
  4AA_3BB_2+ AA_3C^2
  \nonumber\\   \fl&+&   2AA_3A_4B+
  A^2A_2B_1 + AA_1A_2B_1 + 2AA_3BB_1 +
  2A_3A_4BB_1+AA_2A_4C   \nonumber\\
    \fl&+&    A^2A_2B_2+4A_3A_4BB_2+A^2A_2C + AA_3C +
  AA_1A_3C    + AA_1A_2B_2 \, ,\\
 \fl  A'_4&=&2A^2A_2A_3 + 2AA_2^2A_4 + 2AA_2^2B + 2AA_3^2\,B +
  2A_4^3B + 6A_4^2B^2 + 2A_3^2BB_1
  \nonumber\\   \fl&+&   2AA_2^2B_2 +
  4A_3^2BB_2 + 2AA_2A_3C + 2A^2A_4C + A^2C^2 + AA_4C^2\, , \label{eq:b2jednacinaa4}
\\
\fl  B'_1&=&A^2A_1^2 + AA_2^2A_4 + AA_2^2B + AA_3^2B + A_3^2A_4B +
2A^2A_1B_1 + AA_1^2B_1+ 8BB_2^3\nonumber\\  \fl&+&    6B^2B_1^2+
2BB_1^3 + 2AA_2^2B_2 + 8B^2B_1B_2 +
  4BB_1^2B_2+ 8B^2B_2^2 +  8BB_1B_2^2 \, ,
\\
\fl  B'_2&=&A^2A_2A_3 + AA_1A_2A_3 + AA_2^2B + AA_3^2B + A_3^2 A_4B
+ AA_2^2B_1+10BB_1B_2^2 + 6BB_2^3\nonumber\\   \fl&+& 2A^2A_1B_2+
  AA_1^2B_2 + AA_2^2B_2 + 12B^2B_1B_2 +  6BB_1^2B_2 +
    10B^2B_2^2    \,  .
   \label{eq:b2jednacinaB2}
\end{eqnarray}
One can check, by inserting $A_1=A_2=A_3=B_1=B_2=0$ and $A_4=D$,
into equation (\ref{eq:b2jednacinaa4}) for the function $A_4$,
that RG equation (\ref{eq:Db2}) for the function $D$ in the case
of the ASAWs model is recovered. This is not surprising, since it
follows from the definitions of $A_4$ and $D$, and it is certainly
also correct for the $b=3$ fractal equations. However, here we do
not quote the $b=3$ RG equations because they are extremely
intricate. For instance, equation for the parameter $A_1$ has 2753
terms, and it is similar for the remaining $A_i$ and $B_i$
equations.


\section*{References}



\begin{thebibliography}{10}


\bibitem{vc}Vanderzande C, 1998
{\sl Lattice Models of Polymers}, Cambridge: Cambrige University
Press

\bibitem{palissetto}Pelissetto A and Vicari E,  2006 {\sl Phys. Rev.} E
{\bf 73} 051802

\bibitem{stella1} Orlandini E, Seno F and Stella A~L, 2000
 {\sl Phys.~Rev.~Lett.} {\bf 84}   294

\bibitem{stella2} Baiesi M,  Carlon E,  Orlandini E and   Stella A~L,
2001 {{\sl  Phys.~Rev.} E} {\bf 63}   041801

\bibitem{dna1} Marenduzzo D,   Bhattacharjee S~M,  Maritan A,
 Orlandini E and Seno F, 2002 {\sl  Phys.~Rev.~Lett.} {\bf 88} 028102

\bibitem{dna2}Kapri R and Bhattacharjee S~M, 2006
 {\sl J.~Phys.:~Condens.~Matter} {\bf 18}  S215


\bibitem{dna3}Giri D and Kumar S, 2006
{{\sl Phys.~Rev.} E} {\bf 73}   050903(R)

\bibitem{dna4}Kapri R and Bhattacharjee S~M, 2007
 {\sl  Phys.~Rev.~Lett.} {\bf 98}  098101


\bibitem{ks93}Kumar S and Singh Y, 1993
{\sl J.~Phys.~A:~Math.~Gen.} {\bf 26} L987

\bibitem{kspa1}Kumar S and Singh Y,  2001 {\sl Physica} A {\bf 293}
 345

\bibitem{leoni}Leoni P, Vanderzande C and Vandeurzen J,
2001 {\sl  J.~Phys.~A:~Math.~Gen.} {\bf 34} 9777


\bibitem{bata}Mukherji S and Bhattacharjee S M 1995
{\sl Phys. Rev.} E {\bf 52} 1930


\bibitem{haddad2}Haddad T A S,  Andrade R F S and  Salinas S R,
2004 {\sl  J.~Phys.~A:~Math.~Gen.} {\bf 37} 1499

\bibitem{ZivicJSTAT}
\v Zivi\' c I, 2007 {\sl J.~Stat.~Mech.}  P02005

\bibitem{r1} Singh Y, Giri D and Kumar S, 2001 {\sl J.~Phys.~A:~Math.~Gen.} {\bf 34} L67

\bibitem{r2}Rajesh R, Dhar D, Giri D, Kumar S and Singh Y, 2002 {\sl Phys.~Rev.} E {\bf 65}  056124

\bibitem{r3} Owczarek A L,  Rechnitzer A,  Krawczyk J and  Prellberg T, 2007 {\sl J.~Phys.~A:~Math.~Gen.} {\bf 40} 13257

\bibitem{ustenko}Usatenko Z and Sommer J-U, 2007 {\sl J. Stat. Mech.} P10006

\bibitem{dhar78}
 Dhar D, 1978 {\sl J.~Math.~Phys.} {\bf 19}
5

\bibitem{EKM} Elezovi\'c S,  Kne\v zevi\'c M and Milo\v
sevi\'c S, 1987  {\sl J.~Phys.~A:~Math.~Gen.}  {\bf 20} 1215

\bibitem{DharVannimenus}
 Dhar D and Vannimenus J, 1987 {\sl J.~Phys. A: Math. Gen.} {\bf 20}
199

\bibitem{Knezevic} Kne\v zevi\' c M and   Vannimenus V, 1987
{\sl J.~Phys. A: Math. Gen.}  {\bf 20}   L969


\bibitem{EZM} Elezovi\'c-Had\v zi\'c S,  \v Zivi\'c I and Milo\v
sevi\'c S, 2003  {\sl J.~Phys.~A:~Math.~Gen.}  {\bf 36} 1213




\end{thebibliography}
\end{document}